\documentclass[sigconf]{acmart}

\usepackage[inline]{enumitem}
\usepackage{csvsimple}

\AtBeginDocument{%
  \providecommand\BibTeX{{%
    \normalfont B\kern-0.5em{\scshape i\kern-0.25em b}\kern-0.8em\TeX}}}

\setcopyright{acmcopyright}
\copyrightyear{2020}
\acmYear{2020}

\acmConference[DAC '20]{DAC '20: ACM Design Automation Conference}{July 19-23, 2020}{San Francisco, CA}
\acmBooktitle{DAC '20: ACM Design Automation Conference, July 19-23, 2020, San Francisco, CA}

\begin{document}

\title{A Unified Learning Platform for Dynamic Frequency Scaling in Pipelined Processors}

%
\author{Arash Fouman Ajirlou}
\email{afouma2@uic.edu}
\orcid{0002-0870-8930}
\affiliation{%
  \institution{University of Illinois at Chicago}
  \city{Chciago}
  \state{IL, USA}
  \postcode{60607}
}

\author{Inna Partin-Vaisband}
\email{vaisband@uic.edu}
\affiliation{%
  \institution{University of Illinois at Chicago}
  \city{Chciago}
  \state{IL, USA}
  \postcode{60607}
}


\renewcommand{\shortauthors}{Fouman and Partin-Vaisband, et al.}

\begin{abstract}

A machine learning (ML) design framework is proposed for dynamically adjusting clock frequency based on propagation delay of individual instructions. A Random Forest model is trained to classify propagation delays in real-time, utilizing current operation type, current operands, and computation history as ML features. The trained model is implemented in Verilog as an additional pipeline stage within a baseline processor. The modified system is simulated at the gate-level in 45 nm CMOS technology, exhibiting a speed-up of 68\% and energy reduction of 37\% with coarse-grained ML classification. A speed-up of 95\% is demonstrated with finer granularities at additional energy costs. 

\end{abstract}




\maketitle

\section{Introduction}

The primary design goal in computer architecture is to maximize the performance of a system under power, area, temperature, and other application-specific constraints. Heterogeneous nature of VLSI systems and the adverse effect of  process, voltage, and temperature (PVT) variations have raised challenges in meeting timing constraints in modern integrated circuits (ICs). To address these challenges, timing guardbands have constantly been increased, limiting the operational frequency of synchronous digital circuits. On the other hand, the augmented variety of functions in modern processors increases delay imbalance among different signal propagation paths. Bounded by critical paths delay, these systems are traditionally designed with pessimistically slow clock period, yielding underutilized IC performance. Moreover, power efficiency of these underutilized systems also degrades due to the increasing power leakage components. Alternatively, when designed with relaxed timing constraints, integrated systems are prone to functional failures. To simultaneously maintain correct functionality and increase system performance, constraint optimization techniques as well as offline and online models have recently been proposed. Typical approaches include, but are not limited to pipelining, multicore computing, dynamic frequency and voltage scaling (DVFS), and ML driven models\cite{Fields02,Zyuban04, kumar03, rahimi17_slot, Hashemi16, Moghadas18,gepner06,hu04, wu04}.

Propagation delay in a processor is a strong function of the type, input, and output of the current operation, and  computation history \cite{rahimi17_slot}. Intuitively, majority of operations are completed within a small portion of the clock period, as determined by the slowest path in the circuit. Based on path delay distribution, as reported in \cite{Hashemi16}, the operational frequency can be doubled for majority of instructions in a typical program.

While multicore approaches have been demonstrated to partially enhance system performance, the scalability of modern multicore systems is limited by the design complexity of instruction level parallelism and thermal design power constraints.
Thus, speeding the single thread execution is an important cornerstone for enhancing single core performance in modern ICs\cite{Isci06}. This is, therefore, the primary focus of this paper.
The main contributions of this work are as follows:

\begin{enumerate}

\item A systematic flow is proposed and implemented as a unified platform for extracting and processing input features for ML classification of instruction delays.
\item A Random Forest (RF) classifier is trained to classify individual instructions into delay classes based on their type, input operands, and the computation history of the system.
\item A new pipeline stage is integrated within a pipelined MIPS processor.
\item The proposed method is synthesized and verified on LegUp\cite{legup} benchmark suite of programs with Synopsys Design Compiler  in 45 nm CMOS technology node.
\end{enumerate}

\section{Related Works}
Predicting timing violations in a constraint-relaxed system is impractical with deterministic approaches, due to the wide dynamic range of input and output signals (typically 32 or 64 bits), and the variety of instructions in a modern processor. ML based approaches for predicting timing violations of individual instructions have recently been proposed,  which consider the impact of the input operands and  computation history  on timing violations\cite{rahimi17_slot, rahimi17_clim}. While significant for the design process of next generation scalable high performance systems, these approaches have several limitations:\\
\begin{enumerate*} [label={\arabic*}), 
       itemjoin={{\\}}]
\item The output of individual instructions has been considered as a ML feature and exploited in these systems for predicting the timing characteristics of the individual instructions. These predictions are, however, carried out in advance of the instruction execution, when the instruction output is not yet available, limiting the effectiveness of these methods in practical systems. 
\item The modules under the test are studied separately and isolated from other computational and non-computational component (e.g., buffers or multiplexers). Despite the reported high prediction accuracy, the same accuracy results are not expected if the methods are applied to a practical execution unit due to the isolated test environment.

\item Power and timing overheads due to additional hardware are not considered in these papers.
\end{enumerate*}

A bit-level ML based method has been proposed in \cite{rahimi15} for predicting  timing violations with reduced  timing guardbands. While up to 95\% prediction accuracy has been reported with this method, the excessively high, per bit granularity of the ML predictions is expected to exhibit substantial power, area, and timing overheads. These overheads are, however, not evaluated in \cite{rahimi15}.
 Furthermore, a procedure for recovery upon a timing error is not provided and the recovery overheads are also not considered in this work.

As an alternative to fine-grain high-overhead ML implementations, multiple coarse-grain schemes for timing error detection and recovery have been proposed to mitigate the adverse effect of the pessimistic design constraint.
A better-than-worst-case (BTWC) design approach has been introduced in \cite{Hashemi16}. With this approach, the clock period is set to a statistically nominal value (rather than the worst-case propagation delay) and the history of timing erroneous program counters (PCs) is kept in a ternary content-addressable memory (TCAM). The TCAM is exploited for predicting timing violations of the following instructions based on previous observations. Owning to the apparent simplicity of this approach, only bi-state operating conditions (i.e., nominal and worst-case clock frequencies) can be efficiently utilized with this method. Alternatively, the design complexity and system overheads are expected to significantly increase with the increasing number of frequency domains.

A thermal-aware voltage scaling approach has been proposed in \cite{salamat19}. A  voltage selection algorithm is developed and integrated into FPGA synthesis process to dynamically scale the core and block RAM voltages. However, driven by workload and thermal power dissipation, this method supports only coarse-grained voltage scaling.

Predicting program error-rate in timing-speculative processors  has been proposed in \cite{assare19}. A statistical model is developed for predicting dynamic timing slack (DTS) at various pipeline stages. The predicted DTS values are exploited to estimate the timing error-rate in a program. The implementation overheads, and the potential performance or power consumption gains, however, are not reported with this approach.

 ML based methods  for modeling system behavior have also been proposed. For example, in \cite{Moghadas18},  linear regression (LR) has been leveraged for modeling the aging behavior of an embedded processor based on current instruction and its operands, as well as the computation history and overall circuit switching activity. As a result, the timing guardband designed to compensate for aging in digital circuits is effectively reduced given graceful degradation. Reallocation of delay budget is, however, not considered with this method.

ML ICs can exhibit a prohibitively high power consumption and physical size. Also, given specific applications, they may introduce additional delay and increase design complexity. To efficiently exploit ML methods for managing frequency in modern processors, delay, power, and area of ML ICs should be considered.

\section{The proposed ML based frequency scaling}

\begin{figure}[b]
  \includegraphics[width=\linewidth]{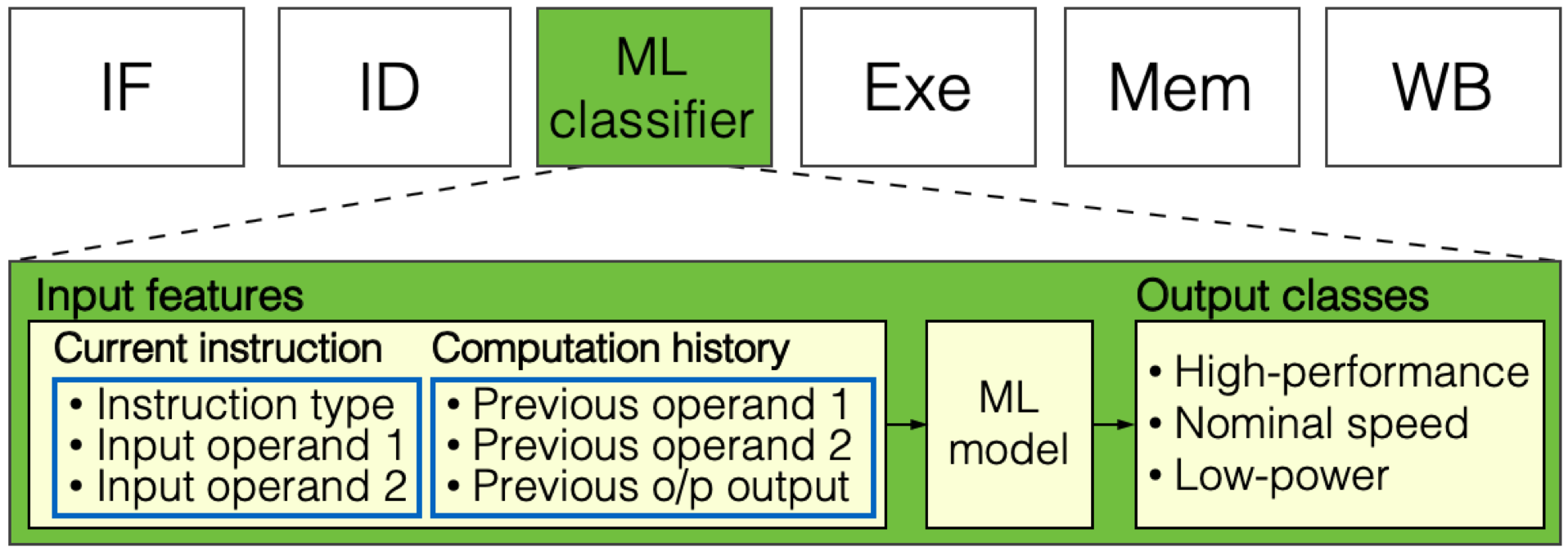}
  \caption{The proposed pipeline with the additional ML stage. In this configuration, six ML features and three delay classes are used.}
  \label{fig:pipeline}
\end{figure}

\begin{figure*}[h]
  \includegraphics[width=\linewidth]{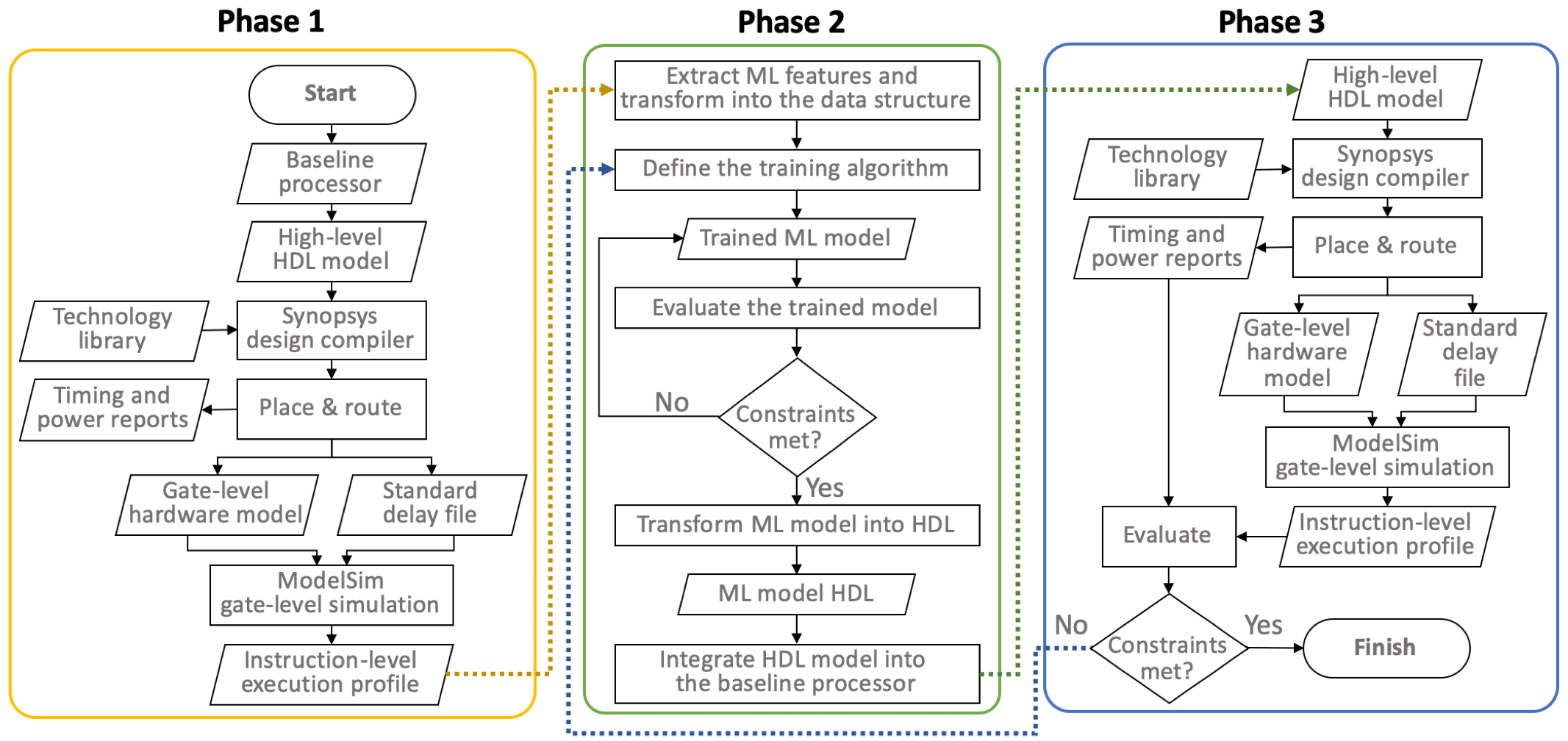}
  \caption{Systematic flow for designing ML predictor within a typical pipelined processor.}
  \label{fig:sysFlow}
\end{figure*}

In this paper, a ML driven design methodology is proposed for pipeline processors. With the proposed method, individual instructions are classified into corresponding propagation delay classes in real-time, and  the clock frequency is accordingly adjusted to narrow down the gap between actual propagation delay and the clock period. The classes are defined by segmenting the worst-case clock period into shorter delay fragments. Each class is characterized by a an operating condition, such as specific supply voltage and clock frequency. The primary design objective is to maximize system performance within the allocated energy budget. 

The frequency of the system is dynamically adjusted in real-time based on the result of instruction delay classification. The overall delay and energy consumption are evaluated with the additional ML components, and both the correct and incorrect predictions. Other control configurations can be defined in a similar manner for different design objectives. In order to evaluate this method, TigerMIPS\cite{TigerMIPS} is utilized as a baseline processor. The ML classifier is designed as an additional pipeline stage within the pipelined MIPS processor, as shown in Fig. \ref{fig:pipeline}. The inputs to the additional ML pipeline stage are the current instruction and its operands, as well as the computation history, as defined by bit-toggled inputs (i.e., current inputs are XORed with the previous inputs), and output of the previous operation. These inputs are utilized as ML features for predicting the delay class of the current instruction based on the trained ML model.  It is important to note that any desired ML model can be trained with this methodology, regardless of its delay, as long as the design complexity and hardware costs of the final system meet the specified constraints. The trained models can be implemented as multiple pipeline stages to meet the timing constraints and maintain the overall system throughput, despite the additional latency  introduced by the ML functions. Also note that the granularity of the output delay classes (e.g., three classes are illustrated in Fig. \ref{fig:pipeline}) can be varied as needed. 

A systematic flow has been developed, implemented, and verified on TigerMIPS with LegUp benchmark suite. The flow comprises three primary phases, as shown in Fig. \ref{fig:sysFlow}. The individual phases are described in the following subsections.

\subsection{Phase 1: Baseline processor synthesis and profiling } 
First, the high-level HDL model of the baseline processor is synthesized into gate-level description model. During this phase, timing information is generated in the IEEE standard delay format (SDF). Based on this information, the gate-level simulation (GLS) is performed and the instruction-level execution profiles are collected. A profile comprises a list of instructions, the fetched or forwarded operands, the output of the operations, and the propagation delays. In addition to the execution profile, post place-and-route (PAR) reports, including timing and power information are collected in this phase.

\subsection{Phase 2: ML training}
In this phase, the gate-level profiles from Phase 1 are parsed and utilized as ML features. The parser also detects and eliminates outliers. The model is trained in Python with Scikit-learn ML library\cite{scikit-learn}. A HDL code (e.g., Verilog in this paper) of the trained model is generated and integrated within the baseline processor as a single (or multiple) pipeline stage(s) between {\it Decode} and {\it Execute} stages (see Fig. \ref{fig:pipeline}).

\subsection{Phase 3: Verification and Evaluation}
Within this phase, the modified high-level HDL model of the system with the ML stage undergoes the synthesis and profiling procedure, as described in Phase 1. To guarantee functional correctness,  the output signal is double-sampled to detect a timing violation, and a timing-erroneous instruction is replayed with the worst-case clock frequency. Similar to the baseline iteration, the post PAR reports are extracted for evaluating the timing and energy characteristics of the system. Finally, the profiling of the modified system is executed during this phase to evaluate the overall speed-up of the system.

To optimize the final solution in terms of the  operational frequency and energy consumption, the proposed flow is executed iteratively, as shown with the feedback in Fig. \ref{fig:sysFlow}. The the clock signal of the pipeline registers is assumed to be near-instantly switched based on the individual classification results, as has been experimentally demonstrated in \cite{Jia19_ISSCC}.

\section{Machine learning background}
To evaluate the efficiency and efficacy of the proposed method, propagation delay classification is investigated with three common ML algorithms: Neural Network (NN), Support Vector Machines (SVMs), and Random Forest (RF). In the following subsections, the primary characteristics of each algorithm are discussed.

\subsection{Neural Networks}

NNs excel in learning complex hidden patterns in large datasets and have shown a particular supremacy in vision and text applications as compared with classical ML algorithms. Following this success, promising results have been shown with NNs in various hardware related applications \cite{NN1,NN2,NN3}. A multi-class NN classifier is designed in this work with a single hidden layer of 20 neurons and ReLu activation function in Scikit-learn ML framework. The network is trained using backpropagation algorithm for 200 epochs until convergence with quasi-newton optimizer. As a general rule, learning capacity of a NN increases with the network complexity (i.e., number of neurons and number of layers). For a NN to be competitive with or outperform classical ML algorithms, large number of neurons and layers is required, significantly increasing the system complexity and hardware overhead of NN based solutions.

\subsection{Support Vector Machines}

SVM classifier learns an optimal hyperplane that separates data samples in feature space with the objective to minimize the classification error. Linear SVM can only learn a linearly-separable decision boundary. Alternatively, to learn complex nonlinear patterns in data, SVM can be combined with a {\it kernel trick} which appropriately transforms the sample features into linearly separable space. In this work, a kernel SVM classifier is designed with the Gaussian kernel. SVM often exhibits excellent performance as compared with other algorithms but suffer from high computational and design complexity, and accordingly high power and area overheads \cite{mohsenin16}.

\subsection{Random Forest}

RF classifier is an ensemble of decision tree classifiers. The input samples are split into multiple sample sets and each decision tree is trained on one training set. The final classification decision for each sample is determined by averaging over the decisions of the trees (i.e., ensembling). RF often benefits from the accuracy, training speed, and interpretability of decision trees, while the ensembling handles overfitting. RF is favorable algorithm in scientific and practical applications \cite{RF, rahimi17_slot}. The computational and hardware complexity of RF is a strong function of the number and depth of decision trees. The depth of the individual trees is determined by the number of features and their correlation. In this work, a RF is trained with low number of shallow trees (i.e., 10 to 100 trees), exhibiting low design complexity and hardware overheads, as demonstrated in Section V.

\begin{table}[t]
\caption{RF, NN, and SVM Configuration and Validation Accuracy.}
\resizebox{\columnwidth}{!}{
\begin{tabular}{||c | c | c | c | c |  c||} 
 \hline
 Algorithm    &\# of classes& Hyperparameter     & Accuracy  & F1-score  & Speed-up  \\ [0.5ex] 
 \hline\hline
 		& C = 2   & \#estimators=10   &  98\%  &  98\%  & 68\%\\
 RF		& C = 3   & \#estimators=100 &  94.1\%  &  94\%  & 70\%\\
  		& C = 4   & \#estimators=100 &  85\%  &  85\%  & 95.3\%\\
\hline		
  		& C = 2   & $h_1 = 20, h_2 = 2$   &  97\%  &  97.4\%  & 68\%\\
 NN		& C = 3   & $h_1 = 20, h_2 = 3$   &  93.1\%  &  92.1\%  & 57\%\\
   		& C = 4   & $h_1 = 20, h_2 = 4$   &  85\%  &  85\%  & 81\%\\
\hline		
  		& C = 2   &   			    &  96.7\%  &  97\%  & 68\%\\
 SVM	& C = 3   & kernel = Gaussian&  89\%  &  89.5\%  & 55\%\\
   		& C = 4   & 			    &  84\%  &  84\%  & 77\%\\
 \hline

\end{tabular}
}
\label{table:trainingRes}
\end{table}

\begin{figure*}[h]
  \includegraphics[width=\linewidth]{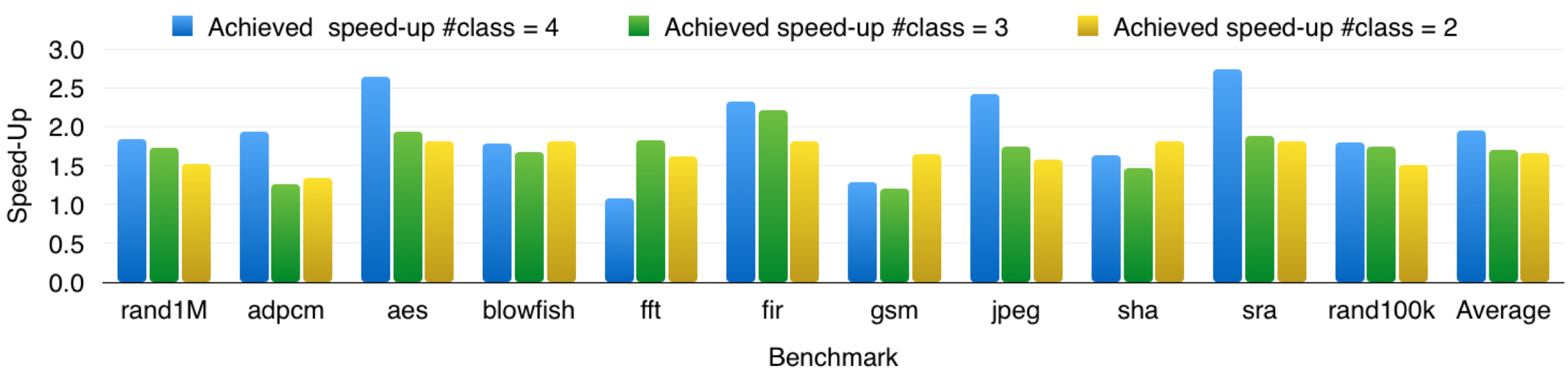}
  \caption{Speed-up with proposed ML framework with two, three, and four delay classes.}
  \label{fig:speed-up}
\end{figure*}

\section{Implementation}
The proposed framework is implemented within TigerMIPS and evaluated  based on LegUp benchmarks. The details of the implementation are described in this section.

\subsection{Unified platform}

A holistic platform is developed to realize the proposed system design methodology, as illustrated in Fig. \ref{fig:sysFlow}. The framework is unified within a shell programming platform supported with several peripheral programs developed in C++ and Python programming languages. The synthesis steps, as described in Fig. \ref{fig:sysFlow}, are sequentially executed from {\it Start} to {\it Finish}. During the first phase, Synopsys Design Compiler is called with the high-level HDL model of the baseline processor.  The profiler triggers are then added to the design and GLS is performed in Modelsim. Phase 2 is triggered upon the completion of the instruction profiling. An external parser program is called to transform the instruction profiles into ML features data structure, and eliminate outliers. The model is trained to classify propagation delays into user-defined number of classes based on a user-specified learning algorithm and delay boundaries. The ML accuracy and estimated speed-up are evaluated upon the training completion. If the design requirements are met, the ML software model is transformed into the high-level HDL code. Otherwise, ML model is retrained with a new algorithm or hyperparameters. Eventually, the HDL code of the ML model is instantiated within the original HDL model of the  baseline processor. Finally, the procedure in Phase 1 is repeated in Phase 3 with the modified processor model, and the overall system performance and overheads are evaluated.

\subsection{Baseline processor}
The proposed framework is demonstrated on a pipelined MIPS processor (i.e., TigerMIPS). In addition to the basic MIPS units, such as Instruction Fetch (IF), Instruction Decode (ID), Execute (Exe), Memory access (Mem), and Write-back (WB), TigerMIPS comprises advanced units, such as,  forwarding unit, branch handling unit, stall logic, and instruction and data caches, which are common in modern pipeline processors.

\subsection{Synthesis and profiling}
The baseline model is synthesized in 45 nm NanGate CMOS technology node with Synopsys Design Compiler. Upon completion of the synthesis, triggers are implemented in Verilog HDL, enabling data and timestamp sampling at the input and output of the execution unit within the MIPS pipeline. The profiling is performed based on GLS with Modelsim simulator.

\subsection{ML algorithm}
The extracted ML features are transformed into a defined data structure and the model is trained with different algorithms (i.e.,  SVM, NN, and RF). The hyperparameters of the ML algorithms are listed in Table \ref{table:trainingRes}.

\subsection{Integration, Verification and Evaluation}
The trained ML model is first validated in Python. The HDL code of the validated ML model is integrated into the baseline processor.  The modified processor is then synthesized and its functionality is verified through GLS. The post PAR reports are utilized to evaluate the modified system with respect to specified design constraints.

\begin{table*}[h]
\caption{ML and System Level Performance with the Proposed Pipelined Classifier. }
\resizebox{\textwidth}{!}{
\begin{tabular}{|| l | c  | c | c | c || c | c | c | c || c | c | c | c ||}
\hline
    \bfseries  Benchmark  & 
    \multicolumn{4}{|c||}{ \bfseries \#class = 4 } & 
    \multicolumn{4}{|c||}{ \bfseries \#class = 3 } &
    \multicolumn{4}{|c||}{ \bfseries \#class = 2 } \\ 
        \cline{2-13} 
    & Accuracy & F1-score & Achieved  speed-up & Ideal  speed-up & Accuracy & F1-weighted & Achieved speed-up & Ideal  speed-up & Accuracy & F1-score & Achieved speed-up & Ideal  speed-up

    \begin{filecontents*}{result.csv}
a,b,c,d,e,f,g,h,i,j,k,l,m
rand1M,0.837,0.849,1.842,2.633,0.925,0.930,1.732,1.933,0.936,0.940,1.528,1.631
adpcm,0.771,0.773,1.936,2.819,0.886,0.884,1.253,2.079,0.945,0.943,1.393,1.715
aes,0.947,0.932,2.635,3.665,0.986,0.985,1.936,2.186,1.000,1.000,1.818,1.818
blowfish,0.844,0.832,1.783,3.435,0.973,0.970,1.676,2.147,1.000,1.000,1.818,1.818
fft,0.750,0.746,1.085,3.256,0.970,0.971,1.833,2.135,0.988,0.986,1.674,1.776
fir,0.950,0.947,2.324,3.708,1.000,1.000,2.219,2.222,1.000,1.000,1.818,1.818
gsm,0.811,0.799,1.288,2.562,0.844,0.836,1.203,1.911,0.989,0.990,1.594,1.673
jpeg,0.868,0.854,2.421,3.078,0.929,0.935,1.742,2.095,0.982,0.981,1.674,1.766
sha,0.698,0.771,1.628,3.585,0.953,0.931,1.469,2.178,1.000,1.000,1.818,1.818
sra,0.977,0.972,2.735,3.793,0.965,0.966,1.883,2.202,1.000,1.000,1.818,1.818
rand100k,0.840,0.851,1.805,2.633,0.922,0.927,1.736,1.941,0.931,0.936,1.528,1.639
\end{filecontents*}
\csvreader[head to column names]{result.csv}{}
    {\\\hline \a &  \b & \c & \d & \e & \f & \g & \h & \i & \j & \k & \l & \m }
 \\\hline   

        \multicolumn{1}{||c|}{    } & 
   	\multicolumn{4}{|c||}{  } & 
	\multicolumn{4}{|c||}{  } &
	\multicolumn{4}{|c||}{ } \\ 
        \multicolumn{1}{||c|}{   Average energy ovehead   } & 
   	\multicolumn{4}{|c||}{  \bfseries 13\% } & 
	\multicolumn{4}{|c||}{ \bfseries 2\% } &
	\multicolumn{4}{|c||}{ \bfseries -37\%} \\ 
        \multicolumn{1}{||c|}{   } & 
    	\multicolumn{4}{|c||}{  } & 
    	\multicolumn{4}{|c||}{   } &
    	\multicolumn{4}{|c||}{ } \\ 
\hline

\end{tabular}
}
\label{table:Res}
\end{table*}

\begin{figure}[h]
  \includegraphics[width=\linewidth]{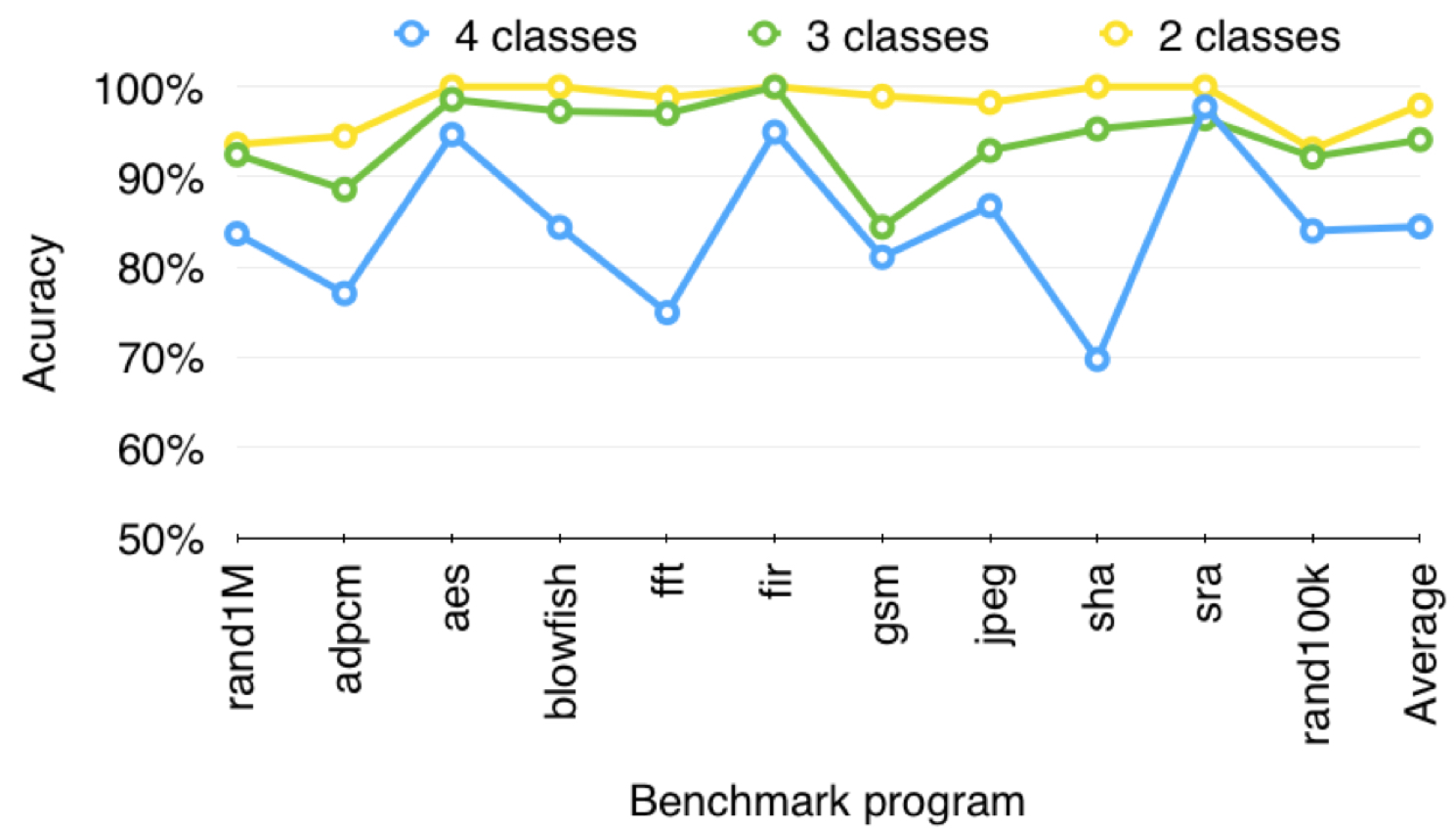}
  \caption{RF classification accuracy in inference on the LegUp benchmark suite with two, three, and four classes. }
  \label{fig:accuracy}
\end{figure}

\section{Experimental Results}
Owing to unique learning characteristics and hardware trade-offs of NN, SVM, and RF models, all these ML models are considered in this paper. Each of these models is trained based on the instruction profiles extracted from a synthetically generated dataset of 3,000 random instructions per class. The boundaries of the individual classes are experimentally determined with respect to the worst-case delay of 4 ns as follows: \{[0.0,2.2],(2.2,4.0]\} for the 2-class configurations, \{[0.0,1.8],(1.8,2.6],(2.6,4.0]\} for the 3-class configurations, and \{[0.0,1.0],(1.0,2.0],(2.0,3.0],(3.0,4.0]\} for the 4-class configurations.

To validate the models, LegUp high-level synthesis benchmark suite coupled with LLMVM compiler toolchain \cite{llvm} is utilized for profiling and verification during GLS. The trained NN, SVM, and RF models with various hyperparameters are validated on the gate-level profiles of the LegUp programs. The average accuracy, F1-score  (a typical accuracy measure which considers precision and recall metrics), and estimated speed-up results are reported in Table \ref{table:trainingRes}.   RF model is preferred in this paper due to its high classification accuracy, higher speed-up, and lower design and hardware complexity as compared with the NN and SVM models.  The trained RF model is tested with nine standard benchmark programs available within the LegUp benchmark suite and two additional synthetically generated benchmarks with one million and 100,000 random instructions. The RF classification accuracy on the test datasets is shown in Fig. \ref{fig:accuracy} for two, three, and four ML delay classes, yielding above 98\% accuracy for majority of the programs with two delay classes. Resultant speed-up for the individual benchmarks is shown in Fig. \ref{fig:speed-up}. A detailed performance and average energy characteristics of the RF model and the modified pipelined processor are listed in Table \ref{table:Res}, including the average ML accuracy, practical reported speed-up (including the misclassification penalty), ideal speed-up (with 100\% classification accuracy), and the energy overhead due to additional ML hardware and classification errors. To account for delay overheads due to the misclassification of a slow instruction into a higher performance class, a replay penalty of four clock cycles is considered within the performance results, as reported in Table \ref{table:Res}. Note that the overall speed-up with four-class configuration is higher than the speed-up with two-class configuration, albeit the higher classification accuracy with two delay classes. Alternatively, higher misclassification rate with four delay classes yields higher replay energy consumption, as listed in Table \ref{table:Res}.

\section{Conclusion}
In this work, an additional ML pipeline stage is proposed for increasing the overall system performance and  temporal resource utilization. This additional stage is designed to classify instructions into propagation delay classes. The system clock frequency is dynamically adjusted based on the individual delay predictions. Pipelining is exploited to mitigate the effect of the ML stage latency on the overall system performance. Practical ML features are extracted based on current instruction and computation history. ML hardware and misclassification power and delay overheads are considered within the reported results. Based on experimental results, up to 95\% performance gain can be achieved with four delay classes at a low energy overhead, and a reduction of 37\% in energy consumption with 68\% gain in performance is practical with two delay classes. A unified shell programing platform with peripheral programs is introduced, yielding a systematic design flow for ML driven pipelined processors.


\begin{thebibliography}{10}

\bibitem{Fields02}
Fields B, Bodík R, Hill MD. Slack: Maximizing performance under technological constraints. InProceedings 29th Annual International Symposium on Computer Architecture 2002 May 25 (pp. 47-58). IEEE.

\bibitem{Zyuban04}
Zyuban V, Brooks D, Srinivasan V, Gschwind M, Bose P, Strenski PN, Emma PG. Integrated analysis of power and performance for pipelined microprocessors. IEEE Transactions on Computers. 2004 Jun 21;53(8):1004-16.


\bibitem{kumar03}
Kumar R, Farkas KI, Jouppi NP, Ranganathan P, Tullsen DM. Single-ISA heterogeneous multi-core architectures: The potential for processor power reduction. InProceedings of the 36th annual IEEE/ACM International Symposium on Microarchitecture 2003 Dec 3 (p. 81). IEEE Computer Society.

\bibitem{rahimi17_slot}
Jiao X, Jiang Y, Rahimi A, Gupta RK. Slot: A supervised learning model to predict dynamic timing errors of functional units. InProceedings of the Conference on Design, Automation \& Test in Europe 2017 Mar 27 (pp. 1183-1188). European Design and Automation Association.

\bibitem{Hashemi16}
Hashemi SH, Ajirlou AF, Soltani M, Navabi Z. Early prediction of timing critical instructions in pipeline processor. In2016 15th Biennial Baltic Electronics Conference (BEC) 2016 Oct 3 (pp. 95-98). IEEE.

\bibitem{Moghadas18}
Moghaddasi I, Fouman A, Salehi ME, Kargahi M. Instruction-level NBTI Stress Estimation and its Application in Runtime Aging Prediction for Embedded Processors. IEEE Transactions on Computer-Aided Design of Integrated Circuits and Systems. 2018 Jun 12.

\bibitem{gepner06} 
Gepner P, Kowalik MF. Multi-core processors: New way to achieve high system performance. InInternational Symposium on Parallel Computing in Electrical Engineering (PARELEC'06) 2006 Sep 13 (pp. 9-13). IEEE.



\bibitem{hu04}
Hu Z, Buyuktosunoglu A, Srinivasan V, Zyuban V, Jacobson H, Bose P. Microarchitectural techniques for power gating of execution units. InProceedings of the 2004 international symposium on Low power electronics and design 2004 Aug 9 (pp. 32-37). ACM.

\bibitem{wu04}
Wu Q, Pedram M, Wu X. Clock-gating and its application to low power design of sequential circuits. IEEE Transactions on Circuits and Systems I: Fundamental Theory and Applications. 2000 Mar;47(3):415-20.

\bibitem{Isci06}
Isci C, Buyuktosunoglu A, Buyuktosunoglu A, Cher CY, Bose P, Martonosi M. An analysis of efficient multi-core global power management policies: Maximizing performance for a given power budget. InProceedings of the 39th annual IEEE/ACM international symposium on microarchitecture 2006 Dec 9 (pp. 347-358). IEEE Computer Society.

\bibitem{legup}
Canis A, Choi J, Aldham M, Zhang V, Kammoona A, Anderson JH, Brown S, Czajkowski T. LegUp: high-level synthesis for FPGA-based processor/accelerator systems. InProceedings of the 19th ACM/SIGDA international symposium on Field programmable gate arrays 2011 Feb 27 (pp. 33-36). ACM.


\bibitem{rahimi17_clim}
Jiao X, Rahimi A, Jiang Y, Wang J, Fatemi H, De Gyvez JP, Gupta RK. Clim: A cross-level workload-aware timing error prediction model for functional units. IEEE Transactions on Computers. 2017 Dec 14;67(6):771-83.


\bibitem{rahimi15}
Jiao X, Rahimi A, Narayanaswamy B, Fatemi H, de Gyvez JP, Gupta RK. Supervised learning based model for predicting variability-induced timing errors. In2015 IEEE 13th International New Circuits and Systems Conference (NEWCAS) 2015 Jun 7 (pp. 1-4). IEEE.

\bibitem{salamat19}
Khaleghi B, Salamat S, Imani M, Rosing T. FPGA Energy Efficiency by Leveraging Thermal Margin. arXiv preprint arXiv:1911.07187. 2019 Nov 17.

\bibitem{assare19}
Assare O, Gupta R. Accurate Estimation of Program Error Rate for Timing-Speculative Processors. InProceedings of the 56th Annual Design Automation Conference 2019 2019 Jun 2 (p. 180). ACM.


\bibitem{TigerMIPS}
Moore, S. and Chadwick, G., 2011. The Tiger ``MIPS'' processor.

\bibitem{scikit-learn}
Pedregosa F, Varoquaux G, Gramfort A, Michel V, Thirion B, Grisel O, Blondel M, Prettenhofer P, Weiss R, Dubourg V, Vanderplas J. Scikit-learn: Machine learning in Python. Journal of machine learning research. 2011;12(Oct):2825-30.

\bibitem{Jia19_ISSCC}
Jia T, Joseph R, Gu J. 19.4 An Adaptive Clock Management Scheme Exploiting Instruction-Based Dynamic Timing Slack for a General-Purpose Graphics Processor Unit with Deep Pipeline and Out-of-Order Execution. In2019 IEEE International Solid-State Circuits Conference-(ISSCC) 2019 Feb 17 (pp. 318-320). IEEE.


\bibitem{NN1}
Yue J, Liu R, Sun W, Yuan Z, Wang Z, Tu YN, Chen YJ, Ren A, Wang Y, Chang MF, Li X. 7.5 A 65nm 0.39-to-140.3 TOPS/W 1-to-12b Unified Neural Network Processor Using Block-Circulant-Enabled Transpose-Domain Acceleration with 8.1× Higher TOPS/mm 2 and 6T HBST-TRAM-Based 2D Data-Reuse Architecture. In2019 IEEE International Solid-State Circuits Conference-(ISSCC) 2019 Feb 17 (pp. 138-140). IEEE.

\bibitem{NN2}
Lee J, Lee J, Han D, Lee J, Park G, Yoo HJ. 7.7 lnpu: A 25.3 tflops/w sparse deep-neural-network learning processor with fine-grained mixed precision of fp8-fp16. In2019 IEEE International Solid-State Circuits Conference-(ISSCC) 2019 Feb 17 (pp. 142-144). IEEE.

\bibitem{NN3}
Lee J, Lee J, Han D, Lee J, Park G, Yoo HJ. 7.7 lnpu: A 25.3 tflops/w sparse deep-neural-network learning processor with fine-grained mixed precision of fp8-fp16. In2019 IEEE International Solid-State Circuits Conference-(ISSCC) 2019 Feb 17 (pp. 142-144). IEEE.

\bibitem{mohsenin16}
Kulkarni A, Pino Y, Mohsenin T. SVM-based real-time hardware Trojan detection for many-core platform. In2016 17th International Symposium on Quality Electronic Design (ISQED) 2016 Mar 15 (pp. 362-367). IEEE.

\bibitem{RF}
Zhang X, Wang W, Zheng X, Ma Y, Wei Y, Li M, Zhang Y. A Clutter Suppression Method Based on SOM-SMOTE Random Forest. In2019 IEEE Radar Conference (RadarConf) 2019 Apr 22 (pp. 1-4). IEEE.

\bibitem{llvm}
Lattner C, Adve V. LLVM: A compilation framework for lifelong program analysis \& transformation. InProceedings of the international symposium on Code generation and optimization: feedback-directed and runtime optimization 2004 Mar 20 (p. 75). IEEE Computer Society.

\end{thebibliography}
\end{document}